\documentclass[floats,aps,showpacs,amssymb,
,secnumarabic%
,tightenlines%
,nofootinbib]{revtex4}
\usepackage{amssymb}
\usepackage{graphics}
\usepackage{graphicx}
\usepackage{amsmath}
\usepackage{amsfonts}
\usepackage{bm}

\begin{document}

\title{Cosmology with bulk viscosity and the gravitino problem}%

\author{L. Buoninfante$^a$ and G. Lambiase$^{a,b}$}
\affiliation{$^a$Dipartimento di Fisica "E.R. Caianiello" Universit\'a di Salerno, I-84084 Fisciano (Sa), Italy,}
\affiliation{$^b$INFN - Gruppo Collegato di Salerno, Dipartimento di Fisica "E.R. Caianiello" Universit\'a di Salerno, I-84084 Fisciano (Sa) Italy.}
\date{\today}
\def\be{\begin{equation}}
\def\ee{\end{equation}}
\def\al{\alpha}
\def\bea{\begin{eqnarray}}
\def\eea{\end{eqnarray}}

\renewcommand{\theequation}{\thesection.\arabic{equation}}
\begin{abstract}
The gravitino problem is revisited in the framework of cosmological models in which the primordial cosmic matter is described by a relativistic imperfect fluid. Dissipative effects (or bulk viscosity effects) arise owing to the different cooling rates of the fluid components. We show that the effects of the bulk viscosity allow to avoid the late abundance of gravitinos. In particular, for particular values of the parameters characterizing the cosmological model, the gravitino abundance turns out to be weakly depending on the reheating temperature.
\end{abstract}


\maketitle

\section{Introduction}
\setcounter{equation}{0}

Imperfect fluid in cosmology are characterized by the fact that the different components of cosmic fluids are coupled, and having different internal equation of states, their cooling rates turn out to be different as the Universe expands. As a consequence, a deviation of the system from equilibrium occurs. The different cooling rates of the components are responsible for the presence of a bulk viscous pressure of the cosmic medium as a whole. The latter is the only possible dissipative phenomenon for an homogeneous and isotropic Universe\footnote{Let us point out that bulk viscosity effects account for the rapid expansion/compression of fluids that cease to be in thermal equilibrium. Therefore the bulk viscosity gives a measure of the pressure that is necessary for restoring the equilibrium to a expanding/compressed system, a condition that naturally arises in a cosmological expanding Universe.} (see \cite{Zimdahl2001}-\cite{marteensreview} and references therein). 
In particular, in the Friedman-Robertson-Walker (FRW) Universe, the dissipation is described by a scalar quantity (the  bulk viscosity as referred to the thermodynamical approach \cite{marteensreview}).
Bulk viscosity enters Einstein's field equations, altering the thermal histories of (relic) particles, compared with cosmology with perfect fluids, and therefore their abundance\footnote{
A comment is in order. In the standard cosmology, the bulk viscosity effects are absent at high temperatures because, essentially, the bulk viscosity coefficient $\zeta$ is proportional to $\delta\equiv (1-3w)^2$, where $w$ is the adiabatic index (with $w=\partial p/\partial \rho=1/3$ in a Universe radiation dominated). However, such a value of $w$ is relaxed whenever  one considers the interactions among massless (relativistic) particles. These lead to running coupling constants, and hence to a trace anomaly \cite{[38]} $T_\mu^\mu\propto \beta(g) F^{\mu\nu}F_{\mu\nu} \neq 0$. For $SU(N_c)$ gauge theory, characterized by a coupling $g$ and $N_f$ flavors, the effective equation of state turns out to be modified as $1-3 w=\frac{5 g^4}{96 \pi^6}  \frac{[N_c+(5/4) N_f]\,[(11/3)N_c-(2/3)N_f]}{2+(7/2)\,[N_c N_f/(N_c^2-1)]}+{\cal O}(g^5)$, whose numerical value may lie in the range $1-3 w \sim 10^{-2}-10^{-1}$ \cite{[38]}.
To give an estimation of the role of the bulk viscosity, we recall that $\zeta$ is related to the scalar pressure $\Pi$ (that enters directly into cosmological equations, see (\ref{EinsteinEq-FRW})) through the constitutive equations for dissipivative quantities $\Pi=-3\zeta H$. For radiative fluids, i.e. fluids consisting of interacting massless and massive particles, kinetic theory or fluctuation theory arguments allow to derive the dissipative coefficients $\zeta$ in terms of the relaxation time $\tau$: $\zeta=4a_0 T^4 \tau \delta$, where (in units $\hbar=1=c$) $a_0=\pi^2 k_B^2/15\simeq 0.65$ is the radiation constant and $k_B$ the Boltzmann constant (of course the expression for $\zeta$ changes for different fluids \cite{marteensreview}). Comparing the scalar pressure to the radiation energy density $\rho=(\pi^2 g_*)T^4/30$ one gets $\Pi/\rho=24k_B^2 \delta (\tau H)/g_* \lesssim 2.4 \times 10^{-5}$ for $\tau H<1$ and $\delta\simeq 10^{-2}$.
Another interesting possibility to have $w\neq 1/3$ is to consider quantum fluctuations of primordial fields \cite{opher} that lead to $p=(\rho-\langle T_\mu^\mu\rangle)/3$, where $\langle T_\mu^\mu\rangle =k_3\left(\frac{R^2}{3}-R_{\alpha\beta}R^{\alpha\beta}\right)-6k_1\Box R$, $k_{1,3}$ depend on the number of quantum fields (for example, $k_3=\frac{1}{1440\pi^2}(N_0+31 N_1+ 11N_{1/2}/2)\sim 0.07$ for $SU(5)$ model, and similarly for $k_1$), $R=-6({\dot H}+2H^2)$, $R^0_0=-3({\dot H}+H^2)$ and $R^i_i=-({\dot H}+3H^2)\delta^i_i$.}.

The aim of this paper is to explore the implications of cosmology with bulk viscosity in relation to the gravitino problem (see for example \cite{gravitinoPr}). Since gravitino couple to ordinary matter only through the gravitational interaction, it follows that their couplings are Planck suppressed, which implies a (quite) long lifetime
 \[
\tau_{3/2}\sim \frac{M_{Pl}^2}{m^3_{3/2}}\simeq 10^5\left(\frac{1\text{TeV}}{m_{3/2}}\right)^3\mbox{sec}\,.
 \]
Here $m_{3/2}$ is the mass of the particle $\sim 10^2$GeV. Particles with such a long lifetime generate some issues in cosmology since if they decay after the nucleosynthesis, their decay products (which can be gauge bosons and their gaugino partners) would destroy light elements, destroying the successful predictions of Big Bang Nucleosynthesis (BBN). This problem can be avoided putting an upper bound on the reheating temperature. In the framework of GR and from the Boltzamnn equation it turns out that the gravitino abundance $Y_{3/2}=n_{3/2}/s$ (here $s=\displaystyle{\frac{2\pi^2}{45}}\, g_* T^3$) is proportional to the reheating temperature $T_R$ \cite{moroi,bolz,pradler,strumia}
 \begin{equation}\label{Yabundance}
 Y_{3/2} \simeq 10^{-11}\frac{T_R}{10^{10}\text{GeV}}\,.
 \end{equation}
Then by requiring that the abundance (\ref{Yabundance}) remains small for a successful prediction of BBN one gets the constraint on the reheating temperature \cite{olive}
 \begin{equation}\label{upperTR}
T_R \lesssim (10^6 -10^7)\text{GeV} \quad \text{for} \quad m_{3/2} \sim {\cal O}(10^2\text{GeV})\,.
 \end{equation}
In turn this bound opens a serious question for the inflationary models (and for some models of leptogenesis) due to the fact that these models tend to predict a reheating temperature larger than the upper bound (\ref{upperTR}) \cite{linde} (see also \cite{mazumdar}).

The paper is organized as follows. In Section 2 we derive the field equations of the cosmological background in presence of imperfect fluids. The analysis of the gravitino problem in the framework of bulk viscosity cosmology is studied in Section 3. Conclusions are shortly drawn in Section 4.

\section{Imperfect fluid}
\setcounter{equation}{0}

The energy-momentum tensor of imperfect fluids is given by \cite{zimdahl2000}
 \begin{equation}\label{energy-momentum-tensor}
 T_{\alpha\beta}=\rho u_\alpha u_\beta +(p+\Pi)h_{\alpha\beta}+q_\alpha u_\beta+q_\beta u_\alpha +\pi_{\alpha\beta}\,.
 \end{equation}
Here $\Pi$ is the scalar pressure (or the bulk viscous pressure), $h_{\alpha\beta}=g_{\alpha\beta}+u_\alpha u_\beta$ is the projector tensor, $q^\alpha$ is a quantity connected to the flux of the energy, and finally $\pi_{\alpha\beta}$ is the so called anisotropic stress tensor. The quantities $q^\alpha$ and $\pi_{\alpha\beta}$ satisfy the following relations $q^\alpha u_\alpha =0\,, \quad \pi_{\alpha\beta}u^\beta = 0 = \pi^\alpha_{\,\,\,\alpha}$.
By making use of symmetries imposed by the fact that the Universe is isotropic and homogeneous, one gets $q_\alpha =0=\pi_{\alpha\beta}$. The only allowing term is the scalar dissipation $\Pi$ \cite{marteensreview}.
Moreover, one finds that the energy-momentum conservation law $T^{\alpha\beta}_{\,\,\,\,\,;\, \beta}=0$ reads ${\dot \rho}+\Theta (\rho+p+\Pi)=0$.
where the dot stands for the derivative with respect to the cosmic time and $\Theta = u^\alpha_{\,\,\,\, ; \alpha}$.
In an homogeneous and isotropic (spatially flat) Universe, the Einstein equations read \cite{zimdahl2000}
 \begin{equation}\label{EinsteinEq-FRW}
    H^2 = \frac{\kappa \rho}{3}\,, \quad\quad  {\dot H}=-\frac{\kappa}{2}(\rho +p+\Pi)\,,
 \end{equation}
where $H={\dot a}/{a}$ is the Hubble parameter and $\kappa = 8\pi G= 8\pi M_{Pl}^{-2}$ ($M_{Pl} \simeq 1.22 \times 10^{19}$GeV is the Planck mass).

In the radiation dominated era, where the energy density is $\rho=\displaystyle{\frac{\pi^2 g_*}{30}T^4}$ and pressure is $p=\displaystyle{\frac{\rho}{3}}$ ($g_*\sim 10^2$ counts the relativistic degrees of freedom), one finds that Hubble parameter evolves according to the equation \cite{zimdahl2000}
\begin{equation}\label{Hfieldradiation}
  \frac{\ddot H}{H}-2 \frac{{\dot H}^2}{H^2}-6 H^2 c_b^2+\frac{1}{\tau}\left(\frac{\dot H}{H}+2H\right)=0\,,
\end{equation}
where $\tau (=\tau(t))$ is the relaxation time (physically it is interpreted as the mean free time of the relativistic particle) that in general is time dependent, while $c_b^2\equiv \frac{\zeta}{(\rho+p)\tau}$ is the propagation velocity of viscous pulse ($\zeta$ the bulk viscosity coefficient) that may assume values\footnote{The speed of bulk viscosity perturbations $c_b^2$, i.e. non-adiabatic contribution to the speed of sound $v$ in a dissipative fluid without heat flux or shear viscosity (in a viscous medium the sound velocity $v$ propagates with a subluminal velocity), is related to $v$ by the relation $v^2=c_s^2+c_b^2\leqslant 1$, where $c_s=(\partial p/\partial \rho)_S$ ($S$ is the entropy) is the adiabatic contribution to velocity, and the upper limit ensures the causality. Assuming that the pressure and temperature are barotropic, with $p=w\rho=(\gamma-1)\rho$, it then follows $c_s^2=\gamma-1$ and $c_b^2\leqslant 1-c_s^2=2-\gamma$. For a Universe radiation dominated, $\gamma=4/3$, and therefore $c_b^2 \leqslant2/3$. The latter is a general result. In the case of radiative fluids and taking for the trace anomaly corrections $1-3w\sim 10^{-2}$, we get  $c_s^2=\frac{45a_0 \delta}{2\pi^2 g_*}\sim 10^{-5}$. Different results are inferred if one assumes the barotropic forms $\zeta\sim \rho^\varsigma$ and $\tau\sim \rho^\chi$, with $\varsigma$ and $\chi$ constants \cite{marteensreview,chimento,coley}.} $0\leqslant c_b^2\leqslant 2/3$ \cite{Zimdahl1998,Zimdahl1998a,Zimdahl1996,marteensreview,hiscock}. These results refer to the case in which the number of particles is conserved.
Taking into account the particle production, i.e. $\nabla_\mu N^\mu=n\Gamma\neq 0$, it follows that the viscous pressure assumes the form $\Pi=-(\rho+p)\Gamma/\Theta$, with $\Theta=3H$ (obtained for isentropic particle production ${\dot s}=0$), that is entirely determined by the particle production rate. For a radiation dominated era $p=\rho/3$ and $ns=(\rho+p)/T$, so that Hubble parameter evolution is given by
 \begin{equation}\label{Htimeevolution1RD}
    \frac{\ddot H}{H}-\frac{5}{4}\frac{{\dot H}^2}{H^2}+3{\dot H} - 6H^2\left(c_b^2-\frac{1}{2}\right)+\frac{1}{\tau}\left(\frac{\dot H}{H}+2H\right)=0\,.
 \end{equation}
In what follows we shall refer to (\ref{Hfieldradiation}), being results very similar to the case (\ref{Htimeevolution1RD}).


To solve Eq. (\ref{Hfieldradiation}) we look at solutions for $H(t)$ of the form
 \begin{equation}\label{Hgeneral}
   H(t) = H_L \left(\frac{t}{t_L}\right)^\Upsilon\,,
 \end{equation}
where $H_L$, $t_L$ and $\Upsilon$ are undetermined constants. To have an hot early Universe we confine ourselves to the case $\Upsilon<0$. From (\ref{EinsteinEq-FRW}) one infers the relation between the cosmic time and the temperature $T$
 \begin{eqnarray}\label{t-Trelation}
   t &=& t_L C \left(\frac{T}{M_{Pl}}\right)^{\frac{2}{\Upsilon}}\,, \\
   C &\equiv & \left(\frac{8 \pi^3 g_*}{90}\right)^{\frac{1}{2\Upsilon}}\left(\frac{M_{Pl}}{H_L}\right)^{\frac{1}{\Upsilon}}\,,\nonumber
 \end{eqnarray}
which allows to cast the expansion rate in the following form\footnote{Expressions similar to (\ref{HAHGR}) are obtained in different frameworks  \cite{fornengo}. For example, $\nu=2$ in Randall-Sundrum type II brane cosmology \cite{randal}, $\nu=1$ in kination models \cite{kination}, $\nu=-1$ in scalar-tensor cosmology \cite{STcosmology,BD}.}
 \begin{equation}\label{HAHGR}
   H = A(T) H_{GR}\,, \qquad H_{GR}= \frac{1}{2t}\,,
 \end{equation}
where
 \begin{equation}\label{Ageneral}
   A(T)=\eta \left(\frac{T}{T_{ref}}\right)^\nu\,, \quad \nu \equiv \frac{2(\Upsilon+1)}{\Upsilon},
  \end{equation}
  \[
  \eta\equiv 2H_L t_L \left[\left(\frac{4\pi^3 g_*}{45}\right)^{1/2}\frac{T_{ref}^2}{M_{Pl} H_L}\right]^{\frac{\Upsilon+1}{\Upsilon}}\,.
 \]
Here $T_{ref}$ is a reference temperature, that can be fixed for example as the BBN temperature  $T_{BBN}$ ($T_{BBN}\simeq (10^{-2}-10^{-4})$ GeV). Inserting (\ref{Hgeneral}) into (\ref{Hfieldradiation}) one derives the expression for the characteristic relaxation time
 \begin{equation}\label{chartime}
  \tau H = \frac{1+\frac{\Upsilon}{2Ht}}{3c_b^2+\frac{\Upsilon(1+\Upsilon)}{2H^2t^2}}\,.
  \end{equation}
Before to investigate the gravitino problem, we focus on some aspects related to the cosmological model under consideration:
\begin{itemize}
  \item Due to the successful predictions of the BBN, we shall refer to the pre-BBN epoch since it is not directly constrained by cosmological observations. We require that at the instant $t_*$ the Universe starts to evolve according to standard cosmological model, i.e. although the bulk viscosity effects are small during the radiation dominated era, we assume that at $t>t_*$ the adiabatic index is exactly 1/3, so that they vanish. To determine $t_*$, we set $H(t_*)=H_{GR}(t_*)$, that implies
      \begin{equation}\label{t*}
        t_*=t_L\left(\frac{1}{2t_LH_L}\right)^{\frac{1}{1+\Upsilon}}\,, \quad \Upsilon\neq -1\,.
      \end{equation}
      To preserve the BBN predictions, we then require $t_* \lesssim t_{BBN}$, where
      \[
      t_{BBN}\simeq (10^{-2}-10^3)\mbox{sec}\sim
      (10^{22}-10^{27})\mbox{GeV}^{-1}\,.
      \]
  \item To get a first insight of the model (that considerably simplify the Boltzmann equation) we work in the regime $|{\dot T}/T| > H$, where
      \begin{equation}\label{dotT/T}
 \frac{\dot T}{T}=\frac{\Upsilon}{2t}=\frac{\Upsilon}{2C t_L} \left(\frac{T}{M_{Pl}}\right)^{-\frac{2}{\Upsilon}}\,.
       \end{equation}
      The condition $|{\dot T}/T| > H$ occurs for
      \begin{equation}\label{ttilde}
        t< {\tilde t}\,, \qquad {\tilde t}\equiv t_L\left(\frac{|\Upsilon|}{2t_LH_L}\right)^{\frac{1}{1+\Upsilon}}=|\Upsilon|^{\frac{1}{1+\Upsilon}}t_*\,.
      \end{equation}
\end{itemize}


\section{Gravitino problem in cosmology with bulk viscosity}
\setcounter{equation}{0}

As pointed out in the Introduction, gravitino is generated by means of thermal scattering in the primordial plasma.
This occurs during the reheating era after Inflation. To describe the gravitino production one makes use of
the Boltzmann equation for the number density of species in thermal bath. The relevant equation for the
gravitino production is
 \begin{equation}\label{boltzmann1}
   \frac{d n_{3/2}}{dt}+3Hn_{3/2}=\langle \sigma v \rangle n_{\text{rad}}^2\,.
 \end{equation}
Here $n_{3/2, \text{rad}}$ refers to gravitino and relativistic species, while $\langle \ldots  \rangle$
stands for the thermal average of the gravitino cross section $\sigma$ times the relative velocity
of scattering radiation ($v \sim 1$), $\sigma v\sim M_{Pl}^{-2}$. In (\ref{boltzmann1})  the term
$\displaystyle{\frac{m_{3/2}}{\langle E_{3/2}\rangle} \frac{n_{3/2}}{\tau_{3/2}}}$ has been neglected. Here $\displaystyle{\frac{m_{3/2}}{\langle E_{3/2}\rangle}}$ is the average Lorentz factor. Introducing the abundances of the gravitino and of the relativistic particles, $Y_{3/2}=n_{3/2}/s$ and $Y_{\text{rad}}=n_{\text{rad}}/s$, respectively,  the Boltzmann equation (\ref{boltzmann1}) assumes the form
 \begin{equation}\label{boltzmann2}
   \frac{dY_{3/2}}{dT}+\frac{3}{{\dot T}}\left(\frac{{\dot T}}{T}+H\right) Y_{3/2}=\frac{ s \langle \sigma v \rangle}{{\dot T}} Y_{\text{rad}}^2\,.
 \end{equation}
In the regime $|{\dot T/T}|> H$  the Boltzmann equation (\ref{boltzmann2}) reduces to the form
 \begin{equation}\label{boltzmann3}
   \frac{dY_{3/2}}{dT}+\frac{3}{T} Y_{3/2}=\frac{ s \langle \sigma v \rangle}{{\dot T}} Y_{\text{rad}}^2\,,
 \end{equation}
The $Y_{3/2}$-term in (\ref{boltzmann2}) survives because the adiabatic condition is lost in the cosmological model under consideration, contrarily to the standard cosmology.
By integrating from $T_R$ ($\gg {\tilde T}$) to the temperature  ${\tilde T}$, where ${\tilde T} \gtrsim T_{BBN}$ is the temperature corresponding to the instant ${\tilde t}$ defined in (\ref{ttilde}), we find that the general solution to (\ref{boltzmann3}) (with the initial condition $Y_{3/2}(T_R)=0$) is
 \begin{equation}\label{Yfcosmology}
   Y_{3/2}(T={\tilde T}) = {\cal B} \,  \Theta_{\Upsilon}(T_R, {\tilde T})\,,
 \end{equation}
where
 \begin{eqnarray}
   \Theta_{\Upsilon} & \equiv & \left(\frac{{\tilde T}}{M_{Pl}}\right)^\Delta-
   \left(\frac{T_{R}}{{\tilde T}}\right)^3 \left(\frac{T_{R}}{M_{Pl}}\right)^\Delta  \label{Theta}\\
   \Delta &=& 3+\frac{2}{\Upsilon},  \label{Delta}\\
   {\cal B} &\equiv & \alpha_0 \alpha_L\,, \\
  \alpha_0 &\equiv & \frac{2\pi^2g_*}{45(1+3\Upsilon)}\left(\frac{4\pi^3g_*}{45}\right)^{\frac{1}{2\Upsilon}}
   \left[M_{Pl}^2 \langle \sigma v\rangle Y_{\text{rad}}^2\right]\,, \label{alpha0} \\
  \alpha_L &\equiv & (M_{Pl}t_L)\left(\frac{M_{Pl}}{H_L}\right)^{\frac{1}{\Upsilon}}\,, \label{alphaL}
 \end{eqnarray}
with  $\alpha_0 \sim {\cal O}(0-1)$ for values of $\Upsilon$ here considered (the above relations hold for $\Upsilon\neq -1/3$). From Eq. (\ref{t*}) one obtains an expression for $t_L$ given by
 \begin{equation}\label{tL1}
 t_L = t_* (2H_L t_*)^{\frac{1}{\Upsilon}}\,,
 \end{equation}
that inserted into (\ref{alphaL}) yields
 \begin{equation}\label{alphaL1}
 \alpha_L=2^{\frac{1}{\Upsilon}}(M_{Pl} t_*)^{\frac{1+\Upsilon}{\Upsilon}}\,, \quad M_{Pl} t_* \sim 10^{41} \,\,\,
 \text{for}\,\,\, t_*=t_{BBN}\,.
 \end{equation}
Interestingly, $\alpha_L$ is independent on free parameter $t_L$ and $H_L$.

\begin{figure}[h]
  \centering
  \includegraphics[width=3in]{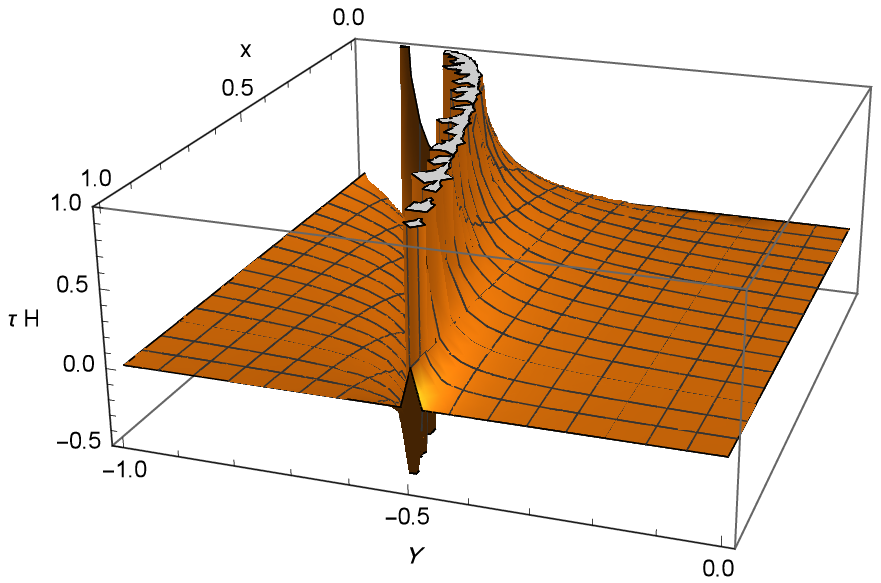}\\
  \includegraphics[width=3.3in]{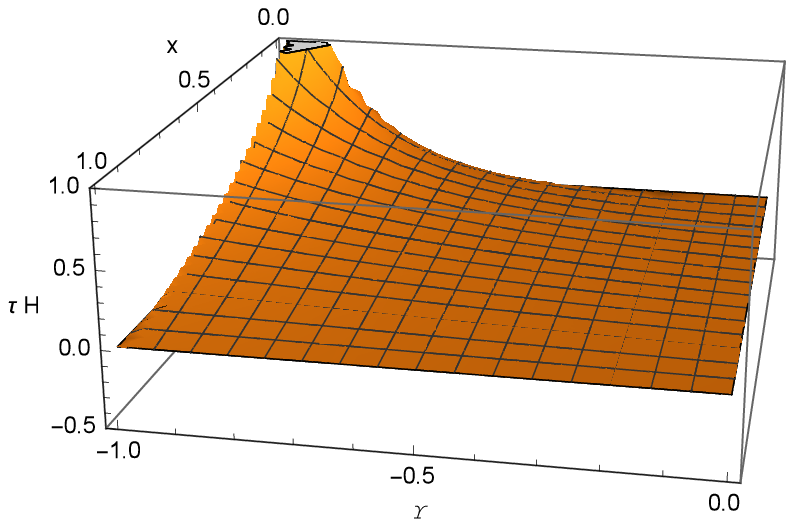}\\
  \caption{$\tau H$ vs $\{\Upsilon, x=t/{\tilde t}\}$ (see Eq. (\ref{tauHgeneral})), with $-1< \Upsilon <0$, $0\lesssim x < 1$, and $c_b^2=2/3$ (upper panel) and $c_b^2=10^{-5}$ (lower panel).}\label{FigtauHgeneral}
\end{figure}

Let us now determine the values of the gravitino abundance $Y_{3/2}$ for different values of $\Upsilon$. First of all, it is simple to show that values of $\Upsilon>0$ yield $Y_{3/2}<0$, which is physically not acceptable. Moreover, Eqs. (\ref{Yfcosmology})-(\ref{alphaL}) and (\ref{alphaL1}) imply that $\alpha_L \ll 1$ for $\Upsilon$ varying in the range $-1< \Upsilon< 0$. Consistently with the cosmological model with bulk viscosity here considered,  we have also to analyze the behaviour of the relaxation time $\tau H$ during the evolution of the Universe. From (\ref{chartime}) it turns out to be given by ($t<{\tilde t}$ in the pre-BBN era)
\begin{equation}\label{tauHgeneral}
  \tau H = \frac{1+ \frac{\Upsilon}{|\Upsilon|}\frac{1}{x^{1+\Upsilon}}}{3c_b^2 + \frac{2 \Upsilon(1+\Upsilon)}{|\Upsilon|^2}\frac{1}{x^{2(1+\Upsilon)}}}\,, \quad
  x = \left(\frac{t}{{\tilde t}}\right)<1\,.
\end{equation}
This function is plotted in Fig. \ref{FigtauHgeneral}, with $c^2_b = 2/3$ and $c^2_b = 10^{-5}$ (see Sect. 2). The parameters $\{\Upsilon, x\}$ have to assume values such that\footnote{
If $\tau$ is of the order of the mean interaction time $t_c=\frac{1}{n\sigma v}$ (here $n$ represents the number density of the target particles with which the given species is interacting, $\sigma$ the interaction cross section, and $v$ the mean relative speed of interacting particles), then the hydrodynamical description requires $\tau H<1$ \cite{marteensreview}.} $0\lesssim \tau H < 1$.

\begin{figure}[h]
  \centering
  \includegraphics[width=3in]{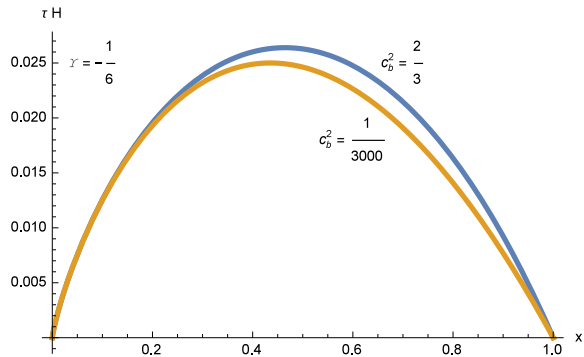}\\
  \caption{$\tau H$ vs $x=t/{\tilde t}$ (see Eq. (\ref{H/HL})), with $0\lesssim x < 1$.}
  \label{relaxtime}
\end{figure}

As arises from (\ref{Yfcosmology})-(\ref{alphaL}), the gravitino abundance is weakly depending on the reheating temperature $T_R$ for a range of values of the parameter $\Upsilon$. Let us set ${\tilde T}=10^{\omega}$GeV. In the case $|\Upsilon|\ll 1$, the gravitino abundance reads
 \[
Y_{3/2}\simeq \alpha_L \left(\frac{M_{Pl}}{{\tilde T}}\right)^\frac{2}{|\Upsilon|}\simeq
\left[2 \times 10^{-2(2+\omega)}\right]^{\frac{1}{|\Upsilon|}}\,,
 \]
which implies $\omega > -2$ in order that $Y_{3/2}\ll 1$ and solve the late abundance gravitino problem.
Taking, for example, $\omega=-1$ and $|\Upsilon|\sim 10^{-2}$ one gets ${\tilde T}\simeq 10^2$MeV and the gravitino abundance $Y_{3/2}({\tilde T})$ is completely negligible. Moreover, in the regime $|\Upsilon|\ll 1$ it follows that the relaxation time is $\tau H\sim \frac{|\Upsilon|x}{2}\ll 1$.
As a specific example, consider $\Upsilon = -1/6$. In such a case the gravitino abundance $Y_{3/2}$ is again weakly depending on the reheating temperature $T_R$. Moreover, one has $t_*=t_L(2t_L H_L)^{-6/5}$, ${\tilde t}\simeq t_*/10$, and the late gravitino abundance turns out to be $Y_{3/2} \sim 10^{-18}\ll 1$. From Eqs. (\ref{Hgeneral}) and (\ref{t*}) one obtains
 \begin{equation}\label{H/HL}
   \tau H=\frac{1-\frac{1}{x^{5/6}}}{3c_b^2-\frac{10}{x^{5/3}}}\,, \quad x=  \left(\frac{t}{{\tilde t}}\right) <1 \,.
 \end{equation}
The function $\tau H$ vs $x$ is plotted in Fig. \ref{relaxtime} for different values of $c_b^2$.  As we can see, the relaxation time is, for all epochs before BBN $t<{\tilde t}< t_*$, smaller than the Hubble time, i.e. $\tau< H^{-1}$, as expected for a fluid description.

\begin{figure}[t]
  \centering
  \includegraphics[width=3.in]{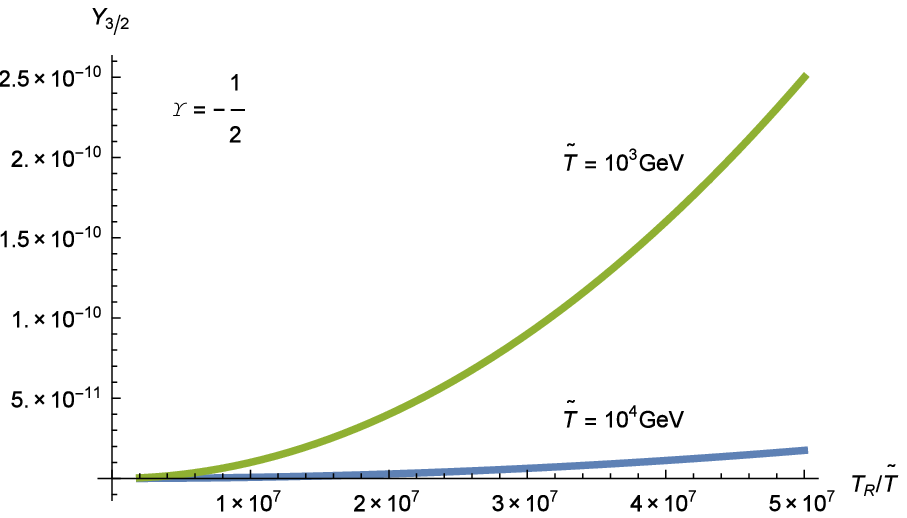}\\
    \includegraphics[width=2.5in]{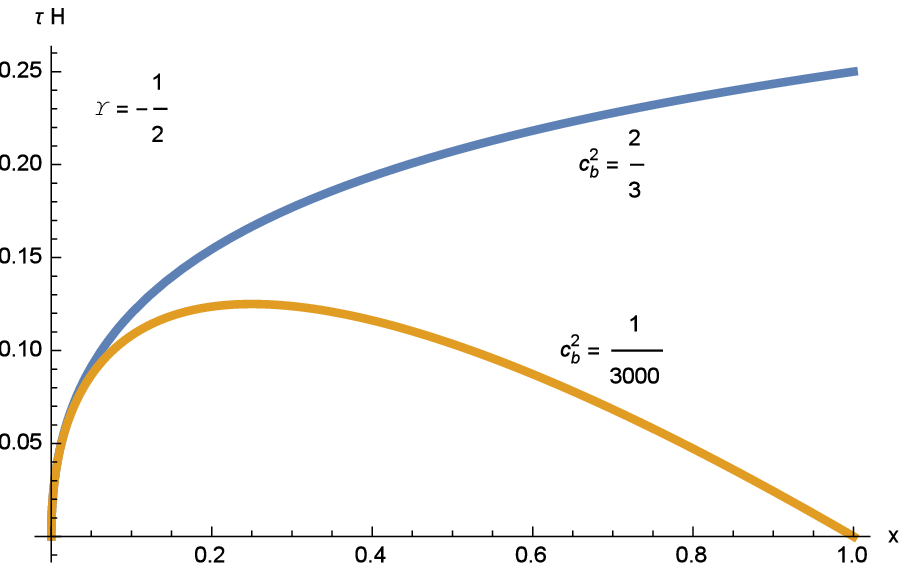}\\
  \caption{Upper panel: $Y_{3/2}$ vs $T_R/{\tilde T}$ (see Eq. (\ref{Yunmezzo})), for different values of ${\tilde T}$ such that $T_R\gg {\tilde T}$. Lower panel: $\tau H$ vs $x=t/{\tilde t}$ (see Eq. (\ref{tauHgeneral})), with $0\lesssim x < 1$.}\label{FigYunmezzo}
\end{figure}

As seen till now, particular values of $\Upsilon$ imply that the late gravitino abundance $Y_{3/2}$ in independent (or weakly depending) on the reheating temperature $T_R$. The latter, however, influences the evolution of $Y_{3/2}$ for increasing values of $\Upsilon$. Consider, for example, the value $\Upsilon=-1/2$. From Eq. (\ref{Yfcosmology}) it follows
 \begin{equation}\label{Yunmezzo}
 Y_{3/2}\simeq 2.5\times 10^{-23}\frac{{\text{GeV}}}{\tilde T}\left(\frac{T_R}{\tilde T}\right)^2\,,
 \end{equation}
while ${\tilde t}\simeq t_*/4$. Notice that $\Theta_{\Upsilon=-1/2}<0$ and $\alpha_0 <0$, so that the two negative signs compensate to give a positive $Y_{3/2}$ (see (\ref{Theta}) and (\ref{alpha0})). The behaviours of $Y_{3/2}$ and $\tau H$ are plotted in Fig. \ref{FigYunmezzo}.


The above analysis refers to $-2/3 < \Upsilon <0$, ($\Delta<0$). Let us analyze now the regime $-1 < \Upsilon \leqslant-2/3$ ($\Delta\geqslant 0$). For $\Upsilon=-2/3$ one obtains ${\tilde t}\simeq 0.3 t_*$, and
  \[
  Y_{3/2}\simeq 1.1\times 10^{-21}\left(\frac{T_R}{\tilde T}\right)^3\,.
   \]
Requiring $Y_{3/2}\lesssim 10^{-10}$ one infers $T_R/{\tilde T}\lesssim 4.6\times 10^{3}$, i.e. to solve the late gravitino overproduction the temperature ${\tilde T}$ must be closer and closer to the reheating temperature $T_R$, which does not seem a favorable scenario. Moreover, the relaxation time turns out to be  $\tau H\gtrsim {\cal O}(1)$, as arises from Fig. \ref{FigtauHgeneral}, making the model not suitable for the solution of the gravitino abundance, at least in the approximation $|{\dot T}/T|>H$.
A similar unfavorable scenario follows also for the cases $-1< \Upsilon <-2/3$ and  $\Upsilon < -1$.

To summarize, the cosmological model with bulk viscosity provides favorable scenarios for solving the late overproduction of gravitino whether the parameter $\Upsilon$ falls down in the range $-2/3 < \Upsilon \lesssim 0$. In this case, in fact, the cosmological evolution of the Universe deviates considerably with respect to the one based on the standard cosmological model (without bulk viscosity effects), as discussed in the previous Section. The range of values $-1 < \Upsilon \leqslant -2/3$, $\Upsilon>0$ and $\Upsilon< -1$, instead, are excluded or partially acceptable, at least in the approximation here considered.

\section{Conclusions}

In this paper, we have reviewed the gravitino problem in a cosmological model in which bulk viscosity effects are taken into account. To avoid the late overproduction of the gravitino by thermal scattering in the primordial plasma, we have exploited the fact that if the cosmic fluid is imperfect then the cosmic evolution of the Universe gets modified as compared to the case of perfect fluids (the expansion rate of the Universe can be written in the form $H=A(T)H_{GR}$, where the factor $A(T)$ accounts for bulk viscosity effects). This affects the Boltzmann equation that describes the time evolution of the gravitino abundance. Moreover, for some choice of the parameters, the gravitino late abundance is weakly depending on reheating temperature. Cosmology with bulk viscosity provides therefore scenarios able to avoid the late overproduction of gravitino. It will be certainly interesting to extend the analysis here studied to the case of non-thermal perturbative gravitino production (see for example \cite{asaka,kawasaki}). This analysis will be faced elsewhere.


\end{document}